# [Book Draft]

# Emotional Interaction between Artificial Companion Agents and the Elderly

By

# Yu Xinjia

Interdisciplinary Graduate School of the

Nanyang Technological University

Singapore

**21 Jan 2016**

# Abstract

Artificial companion agents are defined as hardware or software entities designed to provide companionship to a person. The senior population are facing a special demand for companionship. Artificial companion agents have been demonstrated to be useful in therapy, offering emotional companionship and facilitating socialization. However, there is lack of empirical studies on what the artificial agents should do and how they can communicate with human beings better.

To address these functional research problems, we attempt to establish a model to guide artificial companion designers to meet the emotional needs of the elderly through fulfilling absent roles in their social interactions. We call this model the Role Fulfilling Model. This model aims to use role as a key concept to analyse the demands from the elderly for functionalities from an emotional perspective in artificial companion agent designs and technologies. To evaluate the effectiveness of this model, we proposed a serious game platform named Happily Aging in Place. This game will help us to involve a large scale of senior users through crowdsourcing to test our model and hypothesis.

To improve the emotional communication between artificial companion agents and users, this book draft addresses an important but largely overlooked aspect of affective computing: how to enable companion agents to express mixed emotions with facial expressions? And furthermore, for different users, do individual heterogeneity affects the perception of the same facial expressions? Some preliminary results about gender differences have been found. The perception of facial expressions between different age groups or cultural backgrounds will be held in future study.

This book draft establishes a model to guide artificial companion agent designs based on individual emotional needs. And it also established the mapping between fixed emotion and facial expression of the artificial agents. These are the contributions of this book draft.

# Acknowledgements

I would like to express my sincere thanks and gratitude to my supervisor Prof. Charles Thomas Salmon, my co-supervisor Prof. Cyril Leung, and my mentor Assoc Prof. Miao Chunyan. They guide and support me to find appropriate topic and methods of research. They have shared their experience and spent their time on discussions that had inspired me to go further in my area of interest.

I would like to thank my colleagues and friends at Joint NTU-UBC Research Centre of Excellence in Active Living for the Elderly (LILY) for the support on this book draft work. Their comments and constructive advice are instrumental to the development of this book draft.

# Table of Contents

Table of Contents

Table of Contents

# List of Figures

List of Figures

# List of Tables

# Chapter 1 Introduction

Population aging challenges the world in policy making, health care, science research and other aspects. The majority of the elderly like to age at their own homes not only for the convenience, but also for emotional comfort (Rowles& Chaudhury, 2005; Cristoforetti et al., 2011). Living alone with shrinking social network can negatively impact the physical and psychological wellbeing of the senior citizens (Krause, 1987; Felton & Berry, 1992, Lang & Carstensen, 1994; Takahashi, Tamura, & Tokoro, 1997). Researches have shown beneficial results in solving this problem by using companion agents such as real person, animals and artificial entities (Marcus, 2011). Artificial companion agents have been demonstrated to be useful in therapy (Okita, 2013), offering emotional companionship (Wada, Shibata, & Saito, 2003; Shibata, Kawaguchi, Wada, 2012) and facilitating socialization (Wada, Shibata, Musha, 2008). However, there is lack of empirical studies on what the artificial agents should do and how they can communicate with human beings better.

To address these functional research problems, we attempt to establish a model to guide artificial companion designers to meet the emotional needs of the elderly through fulfilling absent roles in their social interactions. Existing studies found that facial expressions are powerful signals for individuals to understand others' intent in social interactions (Chovil, 1991; Fridlund, 1994). In This book draft, we address an important but largely overlooked aspect of affective computing: how to enable companion agents to express mixed emotions with facial expressions?

This chapter provides an overview of the proposed research. We introduce the background of artificial companion studies and designs. We also highlight the motivations for This book draft from both theoretical and the practical perspectives. The key concepts involved in the proposed research such as emotional support, role absence, and role fulfilling are also defined in this chapter.

Chapter 1 Introduction

## 1.1 Background

Social relationships and social interactions have a positive influence on the wellbeing of the elderly (Weiss, 1977; Cohen & Wills, 1985; Nestmann & Hurrelmann, 1994; Berkman & Glass, 2000). However, for the majority of the senior population, the social relationships around them are incomplete. The studies of artificial companion agents attempt to solve this problem by designing artificial companions to fulfil some of these social roles.

Artificial companion agents can be any kind of hardware or software creations designed to provide companionship to a person. It can be a virtual character in a smart phone app or a physical entity such as a robot. Existing designs focus on several functions of companion effects like cognition assistance (memory, learning, inference, etc.), relieving stress and depression, enhancing social interaction and promoting well-being of the elderly. These functional goals are achieved by the interactions between artificial agents and human beings. These interactions include conversations (verbal, text, tactile, etc.), gestures, and emotional interactions such as facial expressions. Several problems are still unsolved in this area. Firstly, what functions should the companion agents have? Secondly, how should the agents interact with human beings? And are these interactions believable? This book draft focuses on addressing these questions.

## 1.2 Motivation

The motivation for This book draft is to improve the artificial companion agents design to enhance senior citizens' quality of life. By studying the relationships between social role structure, perceived emotion, and wellbeing, This book draft aims to produce a model to guide the device and agent choosing in companion agent design. This book draft aims to support both technology development and offer practical advice to improve communication with the elderly.

Chapter 1 Introduction

## 1.2.1 To guide artificial companion agent design.

For the aim of supporting active living, researchers proposed a variety of technical add-ons such as virtual companion agents, assistive applications and sensors. The problem is the designers lack empirical insights into what the need of the elderly.

The majority of the existing research works related to artificial agents are conducted from the technology designs' perspective. The starting point is often what the researchers believed that the elderly would need. A failed example is surveillance cameras based monitoring systems for the elderly's living environment. Although this technology can observe and record the everyday life and health status of the elderly, and send alarm when emergency happens. However, many seniors regard this as a serious violation to their privacy and refuse to accept such a design. This gap in understanding between designers and elderly end users urgently calls for more large scale and in-depth studies to gain insight into the elderly population.

## 1.2.3 Improve the quality of life for the elderly.

Human beings are social animals. Around each individual, there is a social structure constituted by different roles. For the elderly, they interact with different roles such as family members, friends, social communities and other caregivers and information provides around them. These different roles cannot replace each other as their functions are often different. However, for those who choose to live independently, some of these roles are missing in their daily lives.

The relationships with these roles have a positive influence on an individual's wellbeing (Cantor, 1979; House et al,1988; Mackay et al., 2005). When a role is missing, the older adults tend to feel emptiness and discomfort. Some other negative emotions such as fear and loneliness often follow. In this situation, to fill these missing roles and practice their functions becomes an urgent need technology such as virtual companion agents may be able to fulfil.

## 1.3 Methdology

### 1.3.1 Serious Games

Serious games are designed not purely for entertainment, but also for research, education or other serious purposes. As an emerging method for conducting potentially unobtrusive and large scale studies, serious games are widely used in education, scientific research, health care management, emergency management, and politics studies. A serious game is often a simulation of the real world. For computer science and intelligent agent research, they are usually computer games. During the playing, the players are not only having fun but also performing serious tasks.

In our current work, we incorporated specially designed experiments into a serious game to study individuals' perceived expressions (Yu et al., 2014). The players were asked to express their emotions after each game session in both words and 2D facial expressions. By mapping the words such as "happiness" with the 2D emoticons, we learn how people from different backgrounds relate emotional expressions to composite emotions.

### 1.3.2 Citizen science

Hand (2010) defined citizen science as public participation in scientific research. Citizen science has been demonstrated to be a powerful technique in many fields including systematic collection and analysis of data, development of technology, and testing of natural phenomena (Bhattacharjee, 2005;Hand, 2010).

In our current research, we adopted the technique of implicit human computation to conduct citizen science based studies. We designed a serious game and released it on the Internet to involve volunteers in the experiments. The volunteers access the game by their own computers, then sign up and leave their personal information voluntarily. After logging in, they can play the game. During this process, their performance and behaviour pattern are recorded by the system. This method involved a large number of diverse participants and established an effective dataset.

## 1.4 Organization of the Report

The first section of this book draft has introduced the background and significance of this study. The role filling model is supposed to guide artificial companion design in agent arrangement. And the study of emotional function of role structures can help us to understand the psychological wellbeing of the elderly.

The following Chapter 2 reviewed the prior studies from three aspects. Firstly, the studies on social relationships offer the context and functions of social roles around the elderly. Secondly, the researches of artificial companion agents offered a clear view about the existing agents, devices, their functions and their communication patterns. Then the third part of literature review introduced the models on perceived emotion system. This part gave defined the categories of emotions involved in this studies by the OCC model and discussed how emotions influence behaviours. This chapter established the base of the whole study by defining role, emotion and artificial companion agents.

Chapter 3 introduced the key model of This book draft. The concepts of emotional based role absence and fulfilling are defined in this section. And the model of human-agent interaction based on the roll fulfilling theory. This study believed this roll fulfilling model can improve the future artificial companion design in agent management.

Chapter 4 introduced current results of this study. We suppose the intelligence agents can fill some roles usually played by human beings. To achieve this goal, the agents must have the ability to interact with human beings in emotional ways including expressions. To understand individuals perceived emotion through AI expression, we designed a serious game to collect data. The conclusion of this part offered a base to the both further experiment design and agent-human interaction design.

Chapter 5 as the last part of this book draft introduces some future work. Firstly, the study of facial expression perception based on game platform will involve different age groups. Secondly, we are going to design the proposed role fulfilling model into a new interactive game platform for studying people's self-reported and actual behaviours under its framework.

# Chapter 1 Introduction

Thirdly, we will evaluate the effects of role fulfilling model in guiding companion agent design through large scale user studies based on the game platform.

# Chapter 2 Literature Review

In this chapter, we review the existing works on factors involved in artificial companion agents design including social roles, emotions, and other human factors. We also summarize and categorize the existing research in artificial companion agents designed for living alone elderly. The analysis of their companion functions and interactional patterns are the starting point of This book draft. This chapter also examines the preferences of the elderly to technologies. As most of the works about devices and agents designed for companion purpose focused on technology acceptance, we offer a perspective on the possible factors that may influence the choice of functions and interactional patterns for artificial companion agents.

## 2.1 Social relations of the elderly living alone

### 2.1.1 Social roles around the elderly living alone

Social relationships constitute a major source of personal wellbeing for an individual, especially for the elderly. The social role structure around an older adult represents his/her key sources of care and other support. In 1970s, social support theory started to gain popularity among researchers. It was first used in psychological health studies (Caplan, 1974). Researchers found that social support had the same effect on physical wellbeing by mediating stress and other psychological distresses (Cassel, 1976; Cobb, 1976).

The main social roles around elderly senior citizen can be organized into different levels. Firstly, the closest relationship roles around an individual are his/her family members. Family members share emotions and support each other during their lifetime. They also have a special emotional bond (Moos & Moos, 1976). This is the basis for an individual to reduce isolation, social pressure or fear (Yoo et al., 2014). For those who are suffering from chronic disease or reduced mobility, maintaining good relationship with their families is positively correlated with their wellbeing (Primomo, 2007; Yoo et al., 2014). For the elderly, family members generally include spouse, siblings, children and grandchildren. On this closest level, there are also close old friends. Some research works combined families and friends as one unit in studying the effect of social relationship on individuals (Berkman, 1984; House et al., 1988; Yoo, 2014). This is because some friends involve in the individuals' lives so deep that they are sharing the similar cohesion as their family members.

The second level of social structure consists of community and social groups including neighbours, church groups, and clubs, etc. These communities offer platforms for the individuals to interact with others and make new friends (Wiles al., 2011). People belonging to the same community usually have similar backgrounds. This makes the communities an important source of advices and intellectual stimulation.

The third level consists of care providers including health care workers, doctors, nurses, and information providers (e.g., news media) (Arora et al., 2007). These social roles offer the individual rational information and professional care or advices. These social relationships help an individual maintain a clear understanding of their health status and keep them in touch with the outside world. These supportive social roles can provide confidence and security while reduce the fear of the unknown for the elderly. Since the emotional influence from the third level social relationships is often not as strong as family members or friends, this kind of social relationships is referred to as weak-ties (Yoo, 2014).

In general, family members are closer psychologically to individuals than others in their social network and share the their intimate thoughts and feelings (Manne et al., 2004)

## 2.1.2 Role absence and fulfilment

Extensive researches have shown that social relationships can positively influence the wellbeing of the elderly. A healthy and integral social relationship structure can play a key role in promoting physical and emotional health among the elderly (Berkman, 1984; House et al., 1988; Berkman & Glass, 2000). More frequent social interactions are associated to better self-perceived health status (Cohen & Wills, 1985; House al, 1988; Chang al, 2014). This positive influence works not only on the physical but also on the psychological health (Cohen, 2004). Following their retirement and their children moving out and living on their own, the social networks around older adults are shrinking. This causes some roles in the social role structure around the individual become missing, and negatively influences the older adults' wellbeing, life satisfaction and quality of life (Berkman, 1984; Cohen, 2004).

Researches in computer science and artificial intelligence attempted to fill this role absence in the social structure of the elderly. Computer-mediated social support (CMSS) is one of these attempts. CMSS is a virtual society established within people with similar situation based on social network, private emails and other information exchange channels. This is a useful for fulfilling real social roles, especially for senior patients with chronic disease. CMSS has been demonstrated to improve both the physical and the psychological well-being (Lieberman et al., 2003; Gustafson et al., 2005; Yoo et al., 2014).

The absences of the health care givers' roles around the elderly who live alone can be filled by technology (Lexis al, 2013). This area has attracted significant academic attentions. Glascock and Kutzik introduced the system of "QuietCare" as an intelligent agent to take care of the senior citizens living alone (Clascock & Kutzik, 2012; Lexis al, 2013). The system is based on several sensors in the clients' houses and one data collection unit. It is supposed to provide appropriate and accurate information about the clients' daily living. Systems like this can partially fulfil the absences of the roles of health care givers. For those who need care but do not like strangers in their houses, this type of aging in-place systems can be useful.

Some researchers studied whether nursing staff can take the place of family members of the elderly (Dellasega, 1991; Bowers, 1998; Kellett, 1999; Ryan & Scullion, 2000). These studies believed that even if the staff in nursing homes can offer professional health care, the elderly still like to involve their relatives in their daily care. Bauer and Nay (2003) use the phrase "role loss" to describe the reason of this phenomenon. They believe that although the nurses and doctors can fulfil some functions of daily healthcare, they cannot fulfil the role of family members.

### 2.1.3 Discussions

Social roles around an older individual can be divided in to three levels: family and friends, community and other care givers. Each level has its specific functions and effects on the individual. Close social relationships and effective social support on all the three levels can positively affect individuals' wellbeing in both physical and psychological aspects. When some roles are missing in this structure, other social relationships may be needed to fill their places.

Artificial companion agents also have the ability to play some roles in the social structure around the elderly who are living alone. This is because they can perform multiple functions, such as guiding, offering information, supporting healthcare and accompanying the users.

## 2.2 Artificial companion agents

Artificial companion agents are defined as hardware or software entities designed to provide companionship to a person. It can be a virtual character in a multi-media environment or a physical entity such as a robot. As social animals, human beings tend to rely on companionships from others for cognitive and emotional wellbeing (Trevarthen, 2001). The demand for social communication and companionship is an important behaviour for human beings to form their personality, attitude and life patterns (Buhrmester& Furman, 1987). The senior population are facing a special demand for companionship. On the one hand, they are more susceptible to loneliness, depression and other negative emotions because of their declining physical health and shrinking social circles. On the other hand, since their retirement, losses of loved ones and moving out by their children, companionships from real persons around the elderly are often absent. To fulfil this socio-emotional gap, many researches and businesses attempted to develop artificial companion agents to help the elderly.

### 2.2.1 Existing artificial companion agents

One successful case of the earliest artificial companion agents is the Japanese electronic pet with the name of Tamagotchi. It was a handheld egg-shaped digital pet created by Bandai in 1996. Tamagotchi is a mini computer game based on the simulation of feeding a pet. The pet can communicate with the user by text, status points and facial expressions. Tamagotchi makes the user feel like they are living with a real pet and being needed. It gained great and continued popularity.

Chapter 2 Literature Review

Later, some researchers and businesses turned their attentions to robot pets. The development of these companion agents not only provides the ability of being touched, felt and hugged, but also provides more functions. Kriglstein and Wallner's (2005) team designed HOMIE which is a robot dog with the aim to assist the elderly in daily life. HOMIE can help the user remember their favourite TV shows and channels. It can also remind and help the user with social communication. HOMIE is not only a companion but also an assistance agent. It can show its "emotions" by moving its ears and eyebrows as facial expressions. Although HOMIE is a pet like robot, it can do some simple verbal communication with human beings.

The researches and business development of artificial companion agents showed two different trends. Some of them focus on simulating human beings in terms of functions and communicational patterns, while others do not. Paro as a successful representation of the latter category was developed in laboratory settings in 2003. With a furry appearance like a baby seal, Paro's behaviour is also like a baby or real animal pet (Kidd, 2006). It can understand verbal orders from human beings and respond like some seal sounds or physical gestures. Some researchers showed the therapeutic value of Paro in reducing pain and anxiety of chronic patients (Wada et al, 2003; Shibata, Kawaguchi, &Wada, 2012; Okita, 2013). Some other researchers also found that Paro can improve interactions between human beings (Wada et al, 2008).

One instance of the human-like agents is the TeddyBear which was developed by Fujitsu in 2010. Though the TeddyBear has the appearance of a bear like soft toy, its movements and languages are more like human beings. It can move its body, speak and show facial expressions. Since there is a camera in its nose, it can detect the facial expressions of the user and provide related responses. This is a milestone in the development of the interaction patterns between artificial agents and human beings.

To satisfy more emotional needs and to provide engaging interactions, more companion agents are designed. Teachable agents improve the intergeneration interactions by allowing the elderly to engage in the practice of learning-by-teaching-others (Lim et al., 2013a; Lim et al., 2013b). Curious agents are designed to motivate the elderly to explore new activities by stimulating their curiosity through persuasive techniques (Wu & Miao, 2013). One example of these kinds of agents is Momo which was a computer game based non-person character (Wu & Miao, 2013). It lives in an educational game with the purpose of inspiring the players' curiosity. It can communicate with the users by dialogues based on both text and multiple choices.

## 2.2.2 Discussion

Most of the existing companion agent studies and designs are generated from the designers' perception but not the users'. Most of the companion agents have similar functions such as simulating a real animal pet. However, no research has shown that this companion pattern is the only appropriate one. Are there other needed functions missing? Can other agents with further functions do a better job? What functions are the most appropriate for a specific user? There is a lack of empirical studies to answer these questions and guide the companion agent designs.

During this review, we found that verbal, textual and facial expressions are the most popular interaction functions adopted by existing artificial companion agents. The artificial agents are trained to understand simple languages. This methodology makes the interaction more understandable and believable. However, the interactions based on facial expressions have not been studied in-depth. Can the facial expressions made by the agents be understood by the users? Is there any potential misunderstanding? And furthermore, for different users, do individual heterogeneity affects the perception of the same facial expressions?

These questions generated from the literature review will be the starting point of This book draft.

## 2.3 Models of Emotions

Human beings have various emotions like happiness, anger and surprised. Some of them affect the psychological status of individuals positively. We refer to this kind of emotions as positive emotions. Other emotions which may affect the psychological status of individuals negatively are referred to as negative emotions. There are still some emotions which can be categorised as neutral emotions. In affective computing research, several emotional models from the psychological research field are often used. They are the OCC model (Ortony, Clore

& Collins, 1998), the common sense model (CSM) (Croyle & Barger, 1993), and the plan of behaviour (TPB) model (Ajzen, 1988).

## 2.3.1 The OCC Model

Ortony, Clore and Collins proposed the OCC model to study emotions and the variables affecting their intensities (Ortony, Clore & Collins, 1998). This model categorized 22 emotion types according to the variables that determine their intensities. Most of the emotions identified in their books are organized in pairs of opposite feelings (e.g., "love" and "hate"). The emotions listed in Table 1 serve as the basis of the OCC model.

```
            Joy: (pleased about) a desirable event
       Distress: (displeased about) an undesirable event
      Happy-for: (pleased about) an event presumed to be desirable for someone else
           Pity: (displeased about) an event presumed to be undesirable for someone else
       Gloating: (pleased about) an event presumed to be undesirable for someone else
     Resentment: (displeased about) an event presumed to be desirable for someone else
           Hope: (pleased about) the prospect of a desirable event
           Fear: (displeased about) the prospect of an undesirable event
   Satisfaction: (pleased about) the confirmation of the prospect of a desirable event
 Fears-confirmed: (displeased about) the confirmation of the prospect of an undesirable event
         Relief: (pleased about) the disconfirmation of the prospect of an undesirable event
 Disappointment: (displeased about) the disconfirmation of the prospect of a desirable event
          Pride: (approving of) one's own praiseworthy action
          Shame: (disapproving of) one's own blameworthy action
     Admiration: (approving of) someone else's praiseworthy action
       Reproach: (disapproving of) someone else's blameworthy action
   Gratification: (approving of) one's own praiseworthy action and (being pleased about) the related desirable event
        Remorse: (disapproving of) one's own blameworthy action and (being displeased about) the related undesirable event
      Gratitude: (approving of) someone else's praiseworthy action and (being pleased about) the related desirable event
          Anger: (disapproving of) someone else's blameworthy action and
                 (being displeased about) the related undesirable event
           Love: (liking) an appealing object
           Hate: (disliking) an unappealing object
```

Table 1. model (Ortony, Clore & Collins, 1998)

The OCC model believes emotions are heuristic responses to certain situations. The situations triggering different emotions can be classified into three types: 1) consequences of events, 2) actions of the agents, and 3) aspects of the objects. The structure of this emotional model is shown in Figure 1.

Chapter 2 Literature Review

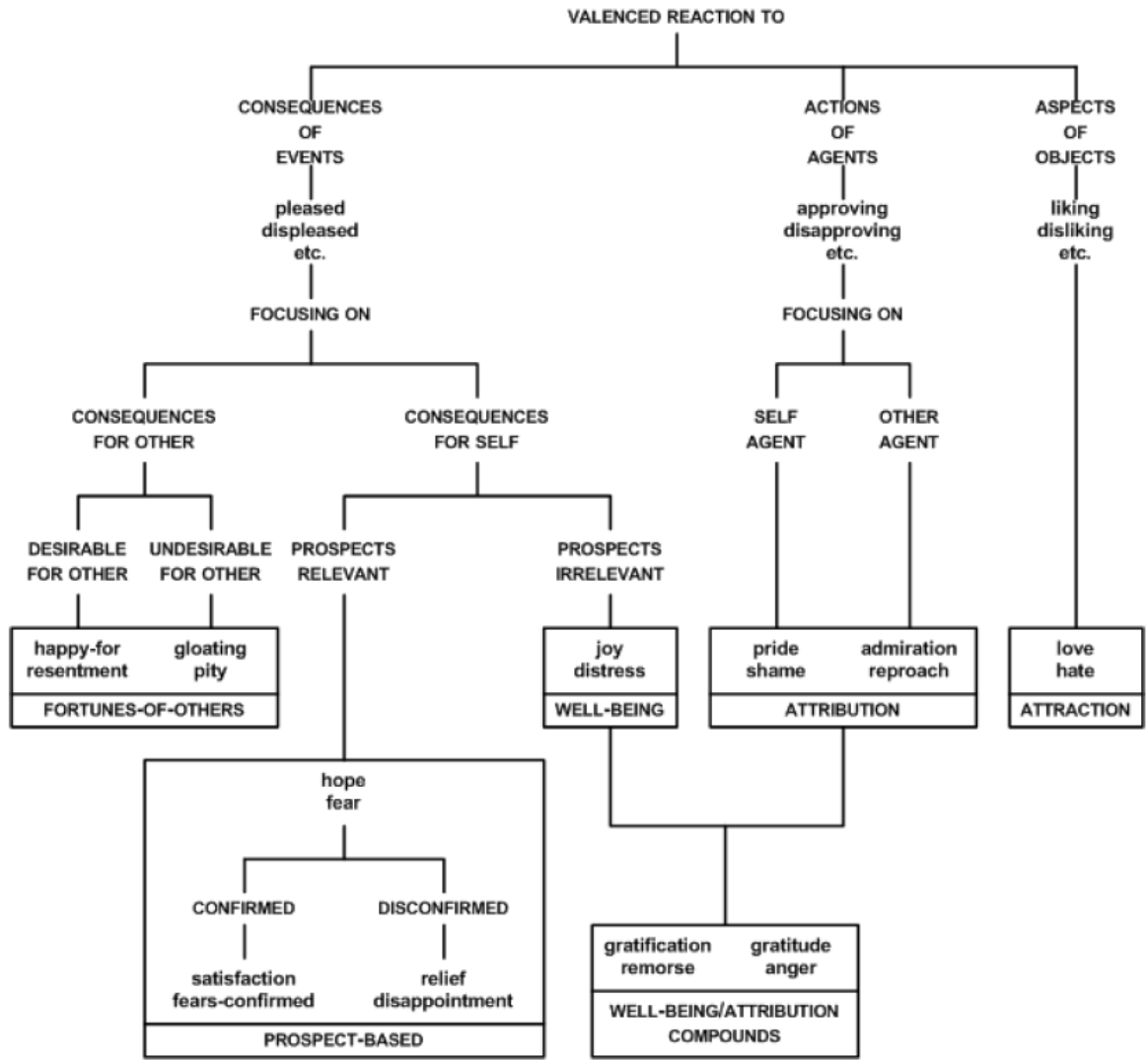

Figure 1 The emotion triggering mechanism in the OCC model (Ortony, Clore & Collins, 1998).

The OCC model is a powerful tool for generating computational emotions for intelligent agents and studying the emotions of human beings because it offers a way to distinguish emotions based on the situations causing them (Lazarus, 1991; Clore &Ortony, 2013). Since emotions are an important part of interactions among human beings, researchers believe it can also affect the interaction between intelligent agents and human beings (Elliott, 1992; Koda, 1996; Bartneck et al, 2008). The OCC model is widely used in generating emotion expressions for artificial intelligent agents (Studdard, 1995; Bondarev, 2002).

## 2.3.2 The Common Sense Model (CSM)

The common sense model (CSM) studies the relationship between cognitive stimuli and coping behaviours (Croyle & Barger, 1993; Leventhal et al., 1998). This model was firstly established by Leventhal, Meyer and Nerenz (1980) to study patients' emotions and behaviour patterns.

CSM believes that when facing stimuli from the outside world, people' response strategy choices are based on their representations of the situation and perceived emotions related to the situation. In other words, how the individuals behave depends what they know and how they feel about it. The representation of external world based on two parts of information gathering. The first source of information is the existing knowledge pool of the individual based on his/her belief system, personal experience, educational and cultural background. The second source includes his/her perceived information about certain time or events from the external social environment (Hagger & Orbell, 2002). Both sources construct the individual's knowledge system about the situation to interpret if it is positive or negative, how will it influence his/her life, and what consequences it will bring (Orbell, al., 1998; Hampson, Glasgow & Stryker, 2000). In respons, they change their usual behaviour patterns by adopting related coping behaviours. The cognitive representation controls the individual's rational behaviour while the cognitive emotion controls his/her emotional behaviour responses. For example, when a pandemic breaks out, the elderly may attempt to seek for medical help, shrink their communicational circle, or change travel plans as a result of the panic caused by the pandemic (Hagger & Orbell, 2001). The logical construction is shown in Figure 2.

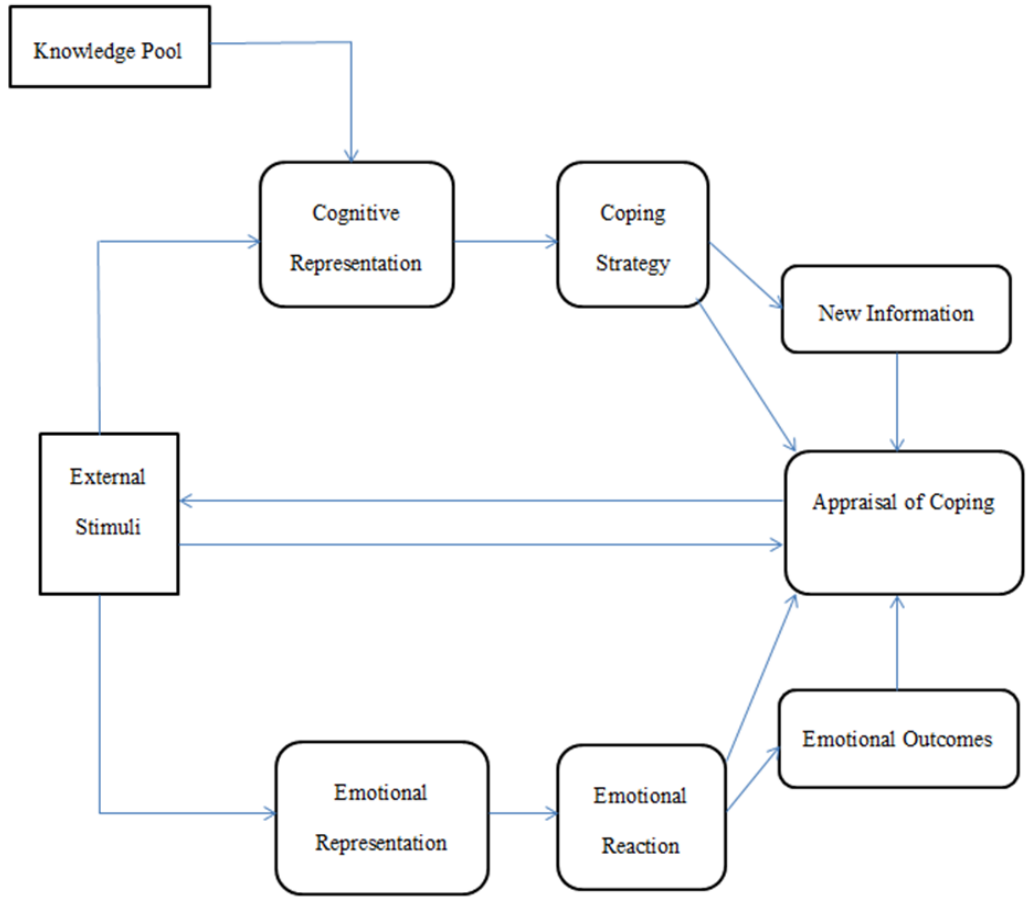

Figure 2 The Common Sense emotion model (Leventhal, Meyer and Nerenz,1980).

### 2.3.3 TPB Model

The Plan of Behaviour (TPB) model integrates individuals' health behaviour intentions (Ajzen, 1988). In the model of TPB, there are three basic constructs: past behaviour, perceived risk of negative outcome, and perceived worry about the negative outcome. The model tries to explain how these three factors affect behavioural intentions by introducing three intermediate variables: attitude toward the behaviour, perceived behavioural control, and normative support for the behaviour.

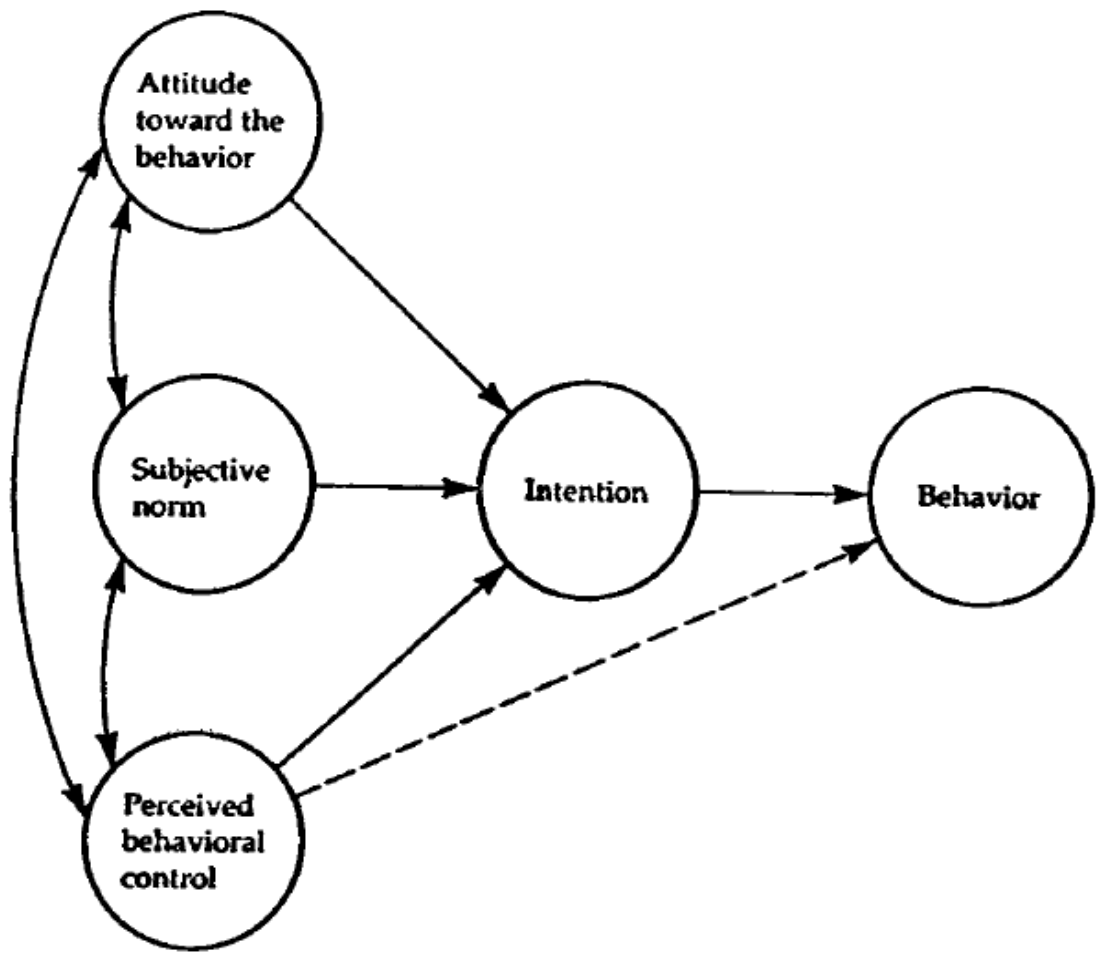

Figure 3 The Plan of Behaviour (TPB) model (Ajzen, 1988).

For example, if the individual needs to make a decision about whether to use a smart watch to measure his /her heart rate, these three factors will determine the final decision according to TPB. The attitude towards the smart watch and the behaviour of wearing it contain the considerations about cost, perceived usefulness and perceived feeling of wearing the watch. The normative support for behaviour is generated from the attitude of the individual's social circle such as his/her families, doctors, and other caregivers. Finally, the perceived behaviour control refers to the cognition of personal ability. In this case, it involves whether the individual believes he/she will wear the watch every day and adjust his/her lifestyle to a healthier way according to the results.

According to TPB, it is possible to change a person's behaviour by changing his/her beliefs about behavioural outcomes and the normative expectations of others. In some cases, when the individual gets more confidence about personal ability in controlling, it is also possible to change his/her behaviour pattern (Ajzen, 1991). TPB has been found effective in changing the behaviour of patients with diabetes, inflammatory bowel disease, and obesity in interactive games (Kharrazi, 2008; 2009; 2010).

Some other researchers also used the model to study how emotions influence behaviours. Schmiege and Klein (2009) tested the relationship between worry (as perceived emotion) and healthy lifestyle behaviours under TPB. They conclude that worry and risk would act as distal predictors of intentions to floss and at least partially mediate proximal behaviours.

### 2.3.4 Discussion

These prior research works provide models to guide the design of emotional interactions between intelligent agents and human beings. The OCC model offers the basis of emotion categories and integrates the path of emotion generation. The CSM model and the TPB model provide tools to connect emotion and behaviour. Till now, we can establish a chain linking external stimuli, emotions, and reactions. This is an important foundation to study the relationship and communication pattern between human beings and intelligent agents, especially under the aging in-place scenario.

# Chapter 3 Role Fulfilling Model

The basic idea to guide aging in-place technology design is to design technologies that can fulfil the missing roles and functions in the everyday life of the elderly. The problems of physical functions are obvious. For a senior citizen who has mobility problem, equipment that can help them move or grasp things can be useful. However, when it comes to the emotional aspect, things are more complicated and less well understood. This book draft aims to establish a model using role as a key concept to analyse the demands from the elderly for functionalities from an emotional perspective in artificial companion agent designs and technologies.

## 3.1 Related work

Social communication can be categorized into three levels (Moulin & Rousseau, 2000):

1) Communication level: activities with the aim of communication maintenance and turn-taking;
2) Conceptual level: concept transformation between agents; and
3) Social level: social relationship management among agents.

The first level supports information exchange among individuals and agents. In this level, the key point affecting communication quality is bi-directional interaction. Human beings' social attribute determines their desires in sending and receiving information. This is just one of the important reasons why computers are more popular than televisions.

On the second level, individuals give order and receive help from other people or agents on the basis of common understanding on related concept. For example, when a customer asks for a cup of coffee, no matter through the waiter or through the ordering system, the only way that this communication will be successful is that the customer and the waiter (or the machine) has the same understanding about the concept of "coffee". This understanding represents the cognitive challenge for artificial agents, especially those designed to help seniors perform activities of daily living.

The third level communications relies on the social relationship structure around an individual. In this structure, the possible behaviours are defined for each role which can be specified through responsibilities, rights, duties, and restrictions (Lewis, 1969). These interaction rules, either implicit or explicit, shape the social relationships around a specific individual. These social roles and the relationships among them compose the social structure around the individual. A contribution from This book draft is to put intelligent agents into the individuals' social structures and accept them to play appropriate social roles.

Hayes-Roth and his team used a metaphor of theatre characters in scenario to describe the social interaction in the real world (Hayes-Roth et al.,1997). They found that social role is a symbol of the behaviour system based on the specific character's status. For example, a lower status role makes the characters humble and submissive to a role with higher status. Roles interact with each other through verbal conversations, gestures and facial expressions. Roles represent a social system about what the individual should do and how others should treat him/her.

The key concepts related to a "role" are role expectation and role awareness. These two concepts connect a single role with its social environment and guide the interactions with the environment and other roles.
Social expectations are generated from the characteristics of the individual performing that role, including his/her social relationships, cultural background, social consensus, etc. It specifies what the society would like the individual to do. Social expectation can work only through the individual's cognition. Role awareness is an important feature of human-human communication. It is what the individual understand about what are expected from him/her, and what kind of person he/she can be (Prendinger & Ishizuka, 2001). The self-awareness is always related to one's skills, ability and background such as age. Self-expectation is a set of personal desire and plan about his/her social status. In another word, self-expectation represents what the individual wants to do and what kind of person he/she wants to be. Social role theory states that the appropriate role of an individual is the intersection of these three factors.

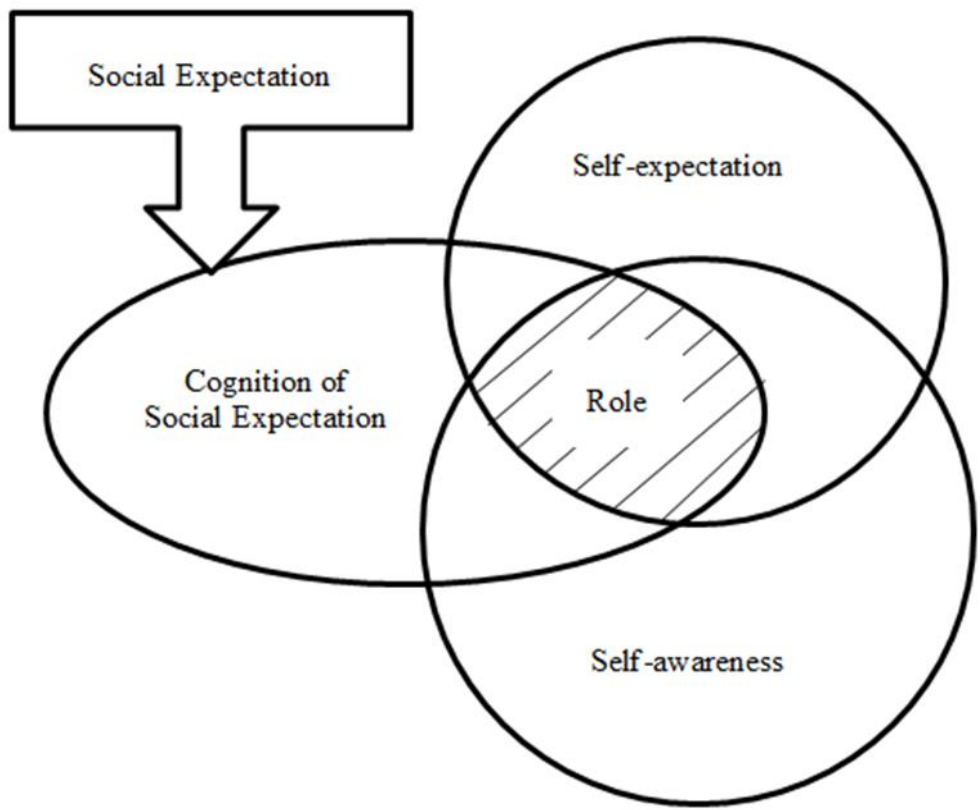

Figure 4 Role expectation and awareness

## 3.2 Roles of Agents in Artificial Companion Agent Applications

If we consider the social relationships around one individual as a small society, his/her families, friends, communities and other care givers all play their own roles within this small society. For the artificial agents, if they were to become an integral part of a user's life, it must play an appropriate role just as the real persons and institutions in this small society. This appropriate role for an aging in-place agent can be discovered by developing the proposed social role model. The key output of this new model is the appropriate role of an agent with a specific set of functions. This new model is a conversation between the potential users and the designers for this prospective agent. The "social expectation" becomes users' expectation. It is generated from the users' need for some specific functions. This expectation can only affect aging in-place design when it is made aware to the designers and researchers. Pre-design role cognition is the designer's idea about the future role the agent is designed to play. To help the agent play that role, this cognition should be related to the expectation of the user. On the other side, what the agent can do are determined by technological limitations and the designers' expectations.

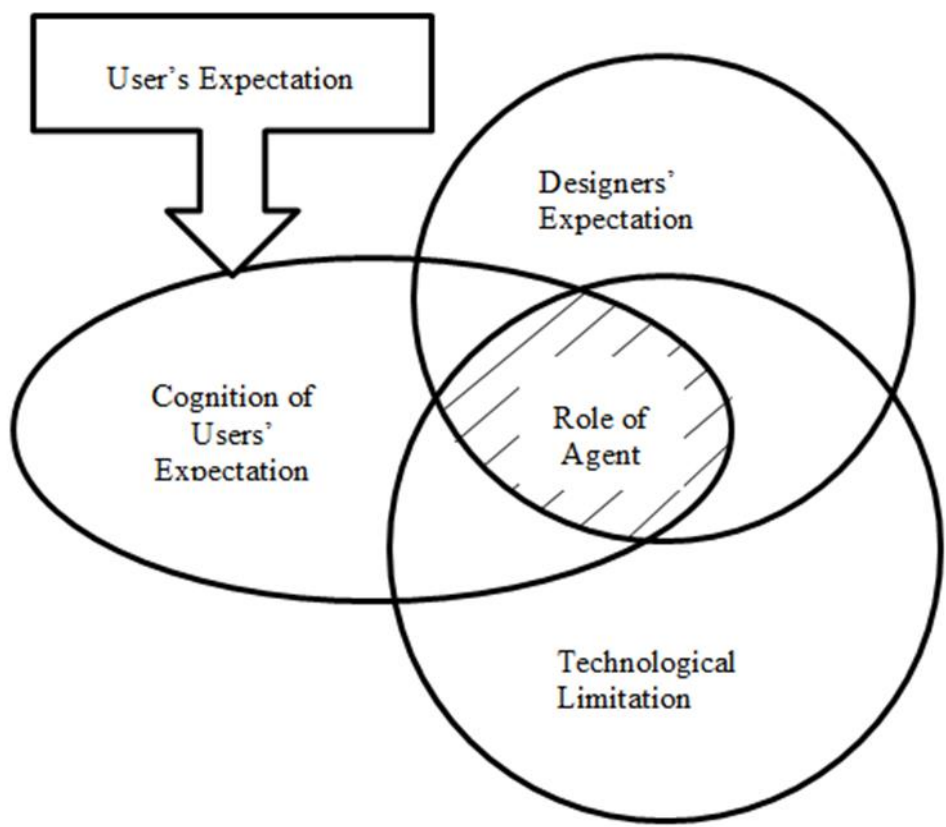

Figure 5 Role expectation and awareness in artificial companion agent design

The designers' expectation and technological limitation determine the possibility of a prospective design. Although designers may be able to generate something to perform a specific function, it may not be needed from the end-users' perspective. For example, a health care agent may be designed to monitor a user's blood pressure. Our expectation of its role is to alert and contact the care givers in case of abnormal blood pressure. The device is working as an observer and information gatherer. Our expectation of this role is to ensure that the elderly living alone can receive timely medical care when having abnormal blood pressure. If the intelligent agents can replace the role as a doctor or nurse to offer medical care anytime, that will be ideal. However, the cost and the technical limitations may not permit such an agent to become a reality.

The proposed model aims to help designers avoid the situations of designing for "no one" or "nobody likes it". It can improve the accuracy and efficiency of aging in-place design. However, how to help designers understand "user's expectation" correctly and efficiently is a major challenge. We can determine the needs for physical assistive devices relatively straight forward, but figuring out the emotional needs of the elderly is difficult. To solve this problem, this study puts forward a model based on the key concept of role – the Role Filling model.

## 3.3 Role Filling Model in Artificial Companion Agent Design

To support aging in-place better, agents cannot simply play the role of tools or toys. They need to integrate into the users' daily lives to play more human-like roles. To achieve this goal, new agent designs should be able to fill the roles missing around elder people who are living alone.

Based on the literature review in Chapter 2, intelligent agents serving a senior citizen can be divided into several categories: daily living activity assistance, work assistance, communication assistance, entertainment, companionship, information provider, and care giver, etc. Nevertheless, some other important roles have been ignored by existing research. Those being taken care of are one of these categories. We call them "beloved" in this study. For most of the elderly, they are the senior generation in their family. During the long past time, they were taking care of their children and the whole family. It is hard for the senior family members to change their roles and attitudes from care givers into receivers. This role missing of the beloved causes two consequences. On one hand, the elderly feel emptiness and useless. This is the reason of the habits of gardening and feeding pets for a lot of senior citizens. They use the plants and pets as beloved to take place of the roles used to be played by their junior family members. On the other hand, some of the elderly never believe in their children's advices. In their mind, children are always children. Receiving care from the children endorses them authority, especially in those family cantered societies as East Asia. That is a serious reason of many family conflicts between generations around the elderly. For the central individual, these roles are playing different roles and support his/her daily life. The user's expectation is determined by whether this individual enjoys the role surrounding including both acceptable and useful feature of the agent. The former one relates to whether the user would like to be surrounded by the agent. The latter one represents whether the agent can help to improve the life quality of the individual. The existing studies about technology acceptance also contained these two aspects. According to the literature review in chapter 2, the factors about concerns of affordability and ease of use, social influence and those related to user's personality are all acceptable factors. And the remaining factors about prospect benefits, need of technology and alternatives are related to usefulness. Both of the two aspects compose the user's expectation of the prospect artificial agents.

Whether a specific agent is welcomed by the elderly can be easily studied by interviews or observations. The question about what kind of agents is expected is hard to answer especially in emotional need aspect. Different agents can influence the users' emotion through multiple channels. According to the OCC model, there are three causes of emotions: consequences of events, actions of agents, and aspects of objects (Ortony et al, 1988).

The role filling model is based on this statement of the OCC model and showed in the following figure.

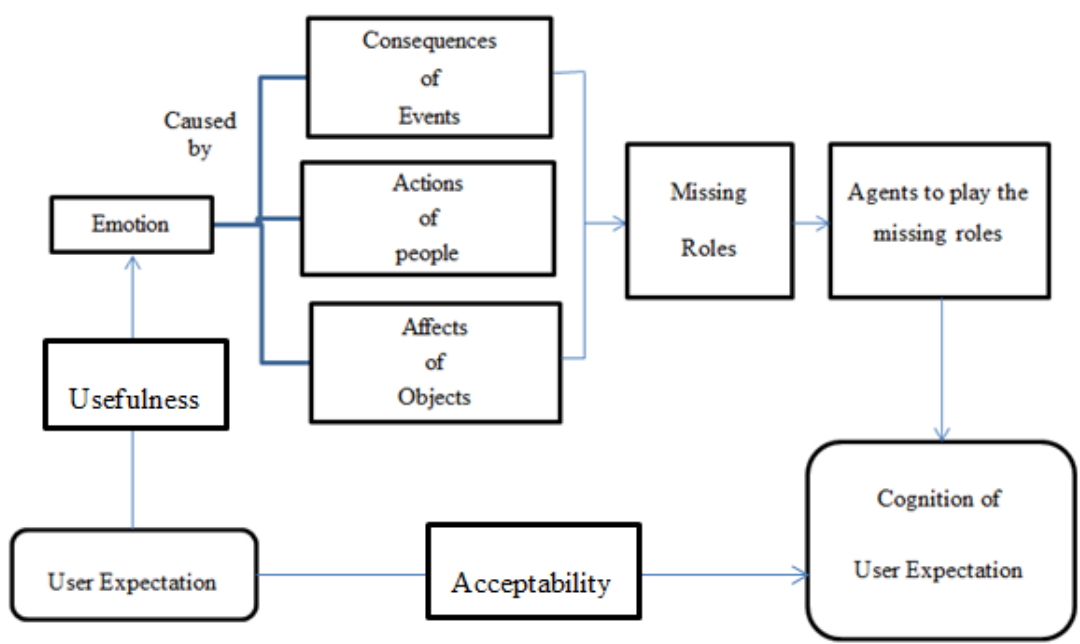

Figure 6 The process from user expectation to the cognition of user expectation

This model aims to offer a effective and efficient briedge for the aging –in-place designers to understand users' expectations. We can use the emotion of fear as an example. When the user is showing siganificant fear, by the explanation of the OCC mdoel showing in figure X, this emotion of fear is caused by the prospective worry about some potential consequences of undisirable events but not by any people or objects. To solve this problem, we need firstly find out the key event. If it is a influenza break out,a better agent should reduce this prospective worry by introducing information as possible consequences of the pandemic. In this case, the information giver is the most important missing role. An appropriate agent design should fill this function. Some others agents as entertainment games with the aim of diverting attention may also work for a while but can not solve the fundemantal problem.

# Chapter 3 Role Fulfilling Model

The roles played by intelligent agents are not replacements of traditional social context but functional supplements. Still us the fear of influenza as an example, to maximize the effort of information giver, we need to let the agents trusted by the central individual to deliver the information. These credible agents can be artificial as health care applications but can also be real agents like families, community leaders, doctors or TV news. The artificial agent in the latter case is a powerful and activated communicational assistance instead of pure information givers.

After combine the role expectation model and the expectation-cognition model together, the role filling model is shown as the following figure.

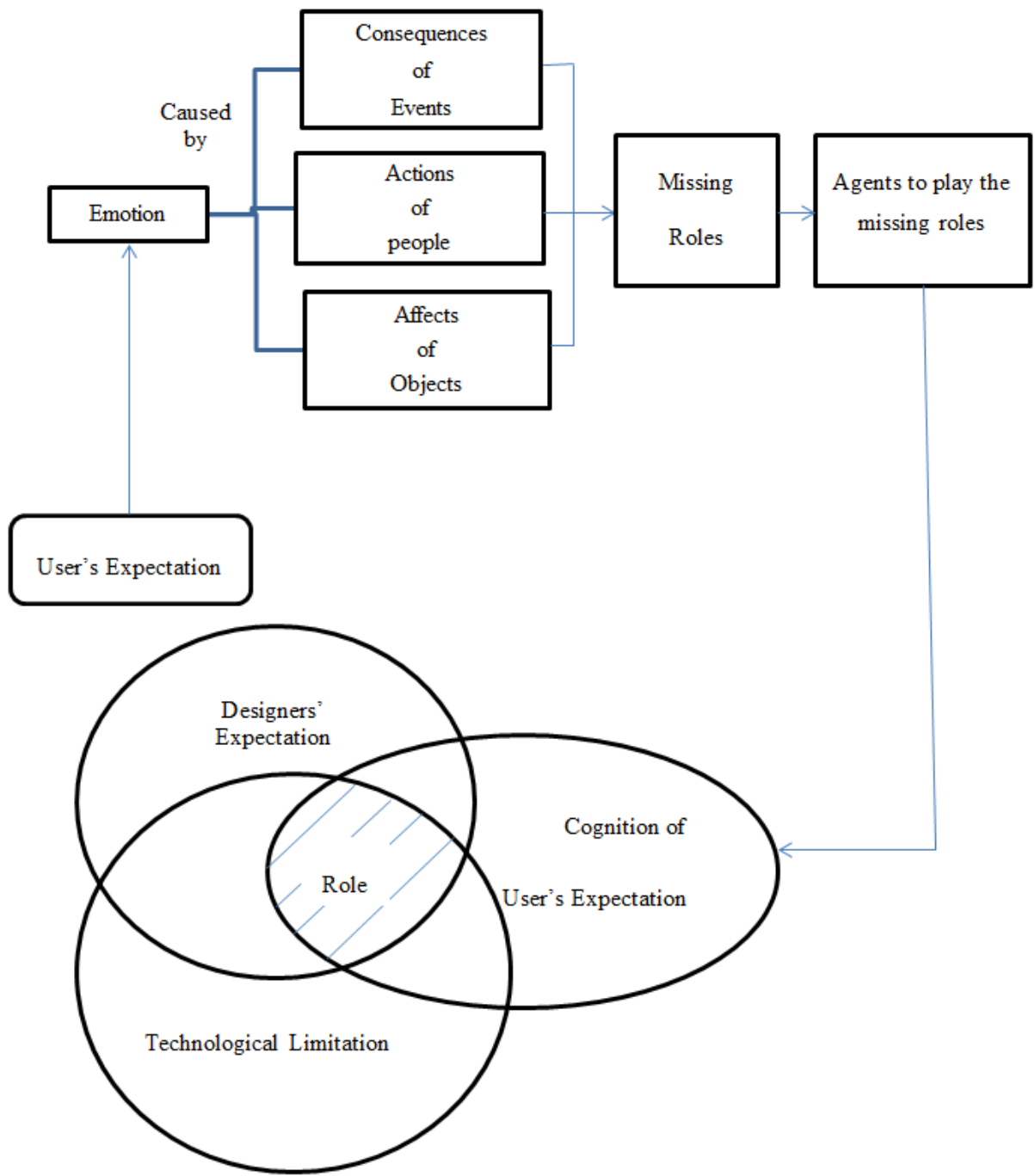

Figure 7 The proposal of artificial companion agent design process based on role fulfilling model

This role filling model aims to guide the aging-in-place design and artificial agent management. It can help us to choose the most appropriate agent for each user's specific need. It can also work in guiding aging-in-place design for a large targeted population. The prior engineering designs start from "identify the problem". And most of the problems just happened at this stage and mislead the whole design process and the failure consequence. The role filling model answers the question about "how to identify the problem". This model based on the detection of emotion of the potential user. The emotion can be recognized and reported in several ways like by word expressions of the user, behaviour detected by sensors and facial expressions. Among these channels, the facial expression is a shortcut and unobtrusive way to get the users emotion. To support this recognition, we need to understand the facial expression communications between human beings and artificial agents. This topic and the related research results will be discussed in Chapter 4.

# Chapter 4.Empirical Studies for Building Emotional Interaction between ACA and the Users

## 4.1. Introduction

Emotion is an important human trait. We use emotions to convey contexts and meanings to others in our daily interactions. Real life emotions are often mixed and involve several simultaneous super-imposed emotions. For embodied agents which need to interact with people (e.g., robot pets, non-player characters in games, virtual assistants in smart phones), two dimensions are crucial for believable human-agent interactions: 1) *Agency*, which refers to the capacity to plan and act; and 2) *Experience*, which refers to the capacity to sense and feel (de Melo et al., 2014). Being able to display proper emotions constitutes an important aspect of the *Experience* dimension.

Past research works have shown that emotion can have a powerful impact on behavior and beliefs (Marsella and Gratch, 2002).Thus, it is important for virtual companions to be able to exhibit proper emotional responses when interacting with human beings in order to build rapport (Moshkina, 2012) and trust (Antos et al., 2011). In This book draft, we address an important but largely overlooked aspect of affective computing: how to express mixed emotions with facial expressions?

Existing research works have proposed computational models for software agents to generate emotions based on appraising its current context. The generated emotions are often in the form of a mixture of multiple emotion types (Dastani and Lorini, 2012).Although a computational approach exists for generating facial expressions from mixed emotions (Albrecht et al., 2005), it is based on complex facial muscle modeling which is computationally intensive and only works for agents with a 3D face embodiment. Due to constraints in cost, computational power and response time, most affective agents still utilize prior drawn facial expressions (e.g., smiley faces) instead of generating them in real-time (Gunes et al., 2011).However, there is a lack of empirical work advising agent designers on how to map mixed emotions to facial expressions.

In order to fill in this gap in affective computing research, we designed and conducted a large-scale user study to construct a mapping from mixed emotions to simple 2D facial expressions. We developed an interactive tool to allow a user to express his/her current feeling in the form of mixture of the Ekman's Six Basic Emotions (i.e., Happiness, Sadness, Excitement, Boredom, Anger, and Surprise) (Ekman, 1992), and select how his/her facial expression would look like in the form of 2D smiley faces. This tool is inserted into a serious game (Yu et al., 2014) as part of the game play.

Using this platform, we conducted user studies following the principles of human computation (Quinn and Bederson, 2011) in 2014 on over 400 university students from China and Singapore. They produced over 2,200 valid responses. In addition to the mapping between mixed emotions and facial expressions, we have also differences in the association between the same facial expressions and mixed emotions between male and female respondents. In this chapter, we discuss the design and limitations of the studies and analyze the results and their implications for agent designers.

## 4.2. Related Work

At the beginning of the 21st Century, artificial intelligence and multi-agent systems researchers started to pay much attention to endowing robots and agents with emotional capabilities. Basic emotions, such as sadness, joy and distress, have been incorporated into the intelligent decision-making processes in response to real-time contexts (Michaud et al., 2000; Chown et al., 2002; Morgado and Gaspar, 2005).

As some robots and agents are designed to act as companions for human beings (be it in the real-world environment or the virtual environment in a game), it is important for the artificial being to exhibit proper emotions in order to gain user acceptance. In (Morgado and Gaspar, 2008), a computational model has been proposed to allow embodied virtual agents to generate and react to emotions when interacting with a human being. Nevertheless, the emotions generated by this model are discrete emotions.

Being a distinctly human trait, it is inevitable for researchers working the field of affective computing to conduct user studies to investigate the properties of emotion in human-agent interactions. In (de Melo et al., 2014), a study of people's interactions with agents with and without emotional responses indicated that artificial emotions can attract people to act favorably towards an agent. This is especially the case if the emotions displayed by the agent are consistent with its past behaviors (Antos et al., 2011). However, these studies still involve only discrete emotions.

Currently, it is technically possible for embodied virtual agents to generate fixed emotions in numerical form. In (Cambria et al., 2012), through affective compression and affective clustering techniques, the authors proposed an approach that can generate the arousal of mixed emotions from interactions in the form of a vector value in a multi-dimensional vector space with each dimension representing a basic emotion. However, a vector representation of mixed emotion cannot be directly interpreted by a human being. It still needs to be converted into facial expressions for instance. Due to the lack of mapping between fixed emotions to facial expressions, the approach in (Cambria et al., 2012) still needs to use the basic emotion with the highest level of arousal and discard the rest when generating facial expressions. With such a mapping, embodied virtual agents remain, to some extent, emotionally autistic when mixed emotions need to be expressed.

In order to establish the mapping between mixed emotions and facial expressions, large-scale user studies are required. An easy to use and scalable tool for people to express their facial expressions in a way that is also suitable for computers to process is needed. In (Broekens and Brinkman, 2013), the *AffectButton* has been proposed for this purpose. It is an interactive self-report mechanism that allows a user to select up to 36 different representative facial expression images by a single mouse click. The images are designed to convey different values along three dimensions: *valence*, *arousal*, and *dominance*. However, this tool has not yet been applied in understanding how the facial expressions related to mixed emotions.

## 4.3. Human Computation based User Studies

The Affect Button has shown wide ranging applicability in domains such as online preference elicitation, and emotion words labeling, etc. It has the potential to be used as a basis for expressing mixed emotions. In this study, we leverage the Affect Button to build a tool for people to explicitly express mixed emotions and the corresponding facial expressions as a component in a serious game. In this section, we discuss the design of the emotion expression study, and the characteristics of the study participants. We embedded this study into a serious game. After playing the game, the participants are asked to choose the closest statement of their emotion in both text and facial expression.

### 4.3.1. Design of the Emotion Expression Component

We aim to design the mixed emotion self-reporting tool as an unobtrusive part of the game play so that people will not feel that they are being asked for about their feelings, but rather see it as a tool for them to be expressive in the game. Such a design philosophy stems from the principles of human computation where the opinions, intelligence, and/or productivity of human beings are harnessed to accomplish tasks which are not suitable for automation (Doan et al., 2011). According to (Quinn and Bederson, 2011), two important characteristics constitute a human computation system:

1) The problems fit the general paradigm of computation, and
2) The human participants are guided by the computational system or process.

The most popular example of human computation is the reCAPTCHA system (von Ahn et al., 2008). A person spends a few seconds recognizing distorted alphanumeric characters from pre-computer age publications when logging into protected online services. Over 100 million CAPTCHAs are recognized every day. By using it as an essential step in protecting online services from abuse, CAPTCHA serves the dual purpose of security while harnessing the fragments of human intelligence and productivity to accomplish the challenging task of digitizing pre-computer age publications.

The reCAPTCHA system is regarded as an implicit human computation approach (Quinn and Bederson, 2011).The participants derive direct utility from the system (i.e., gaining access to password protected services in reCAPTHCA's case) by contributing their manpower, but are unaware of how their contributions are used by the system (i.e., digitizing pre-computer age publications). We take a similar approach when design the proposed mixed emotion self-reporting tool.

The process for a player to complete a level in the AM game appears to the players are the same to most game playing experiences. It is as follows: 1) Before the start of a game session, the specific characteristics and objectives of the level is explained to the player; 2) During the playing session, the player makes task delegation decisions by clicking on each of the PAs under his/her control and observing their responses; 3) After the completion of the game level, the player's score and his/her relative performance compared to the AI competitor is displayed; and 4) At the end of a game session, we embed the proposed fixed emotion self-report tool to enable the player to be expressive about his/her playing experience.

Figure 8. The game interface for the proposed mixed emotion self-report tool.

The design of the proposed tool is shown in the bottom half of Figure 3. It provides an opportunity for the player to reflect on his/her most recent playing experience. The tool consists of two parts. The first part asks the player how he/she feels after playing the game. The play can respond by selecting any one, or a mixture, of the six basic emotions proposed in (Ekman, 1992). The arousal level of each of the basic emotions is presented to the player as a five-star scale. In the background, the value is captured as an 11-point Likert scale

Chapter 4.Empirical Studies for Building Emotional Interaction between ACA and the Users

(Likert, 1932) value (0: no arousal; 10: maximum arousal). The player can simply click on the 5-star scales to indicate he/her mixed emotion.

The second part of the tool asks the player which of the smiley faces best fits how he/she feels at the moment. The images are from one of the most recent and widely used tools in affective computing research - the Affect Button (Broekens and Brickman, 2013). The Affect Button contains 36 images of different facial expressions and 1 image of no expression as shown in Figure 4. The "+" signs on the images in Figure 4 indicates the mouse click position that will trigger the corresponding image to be displayed.

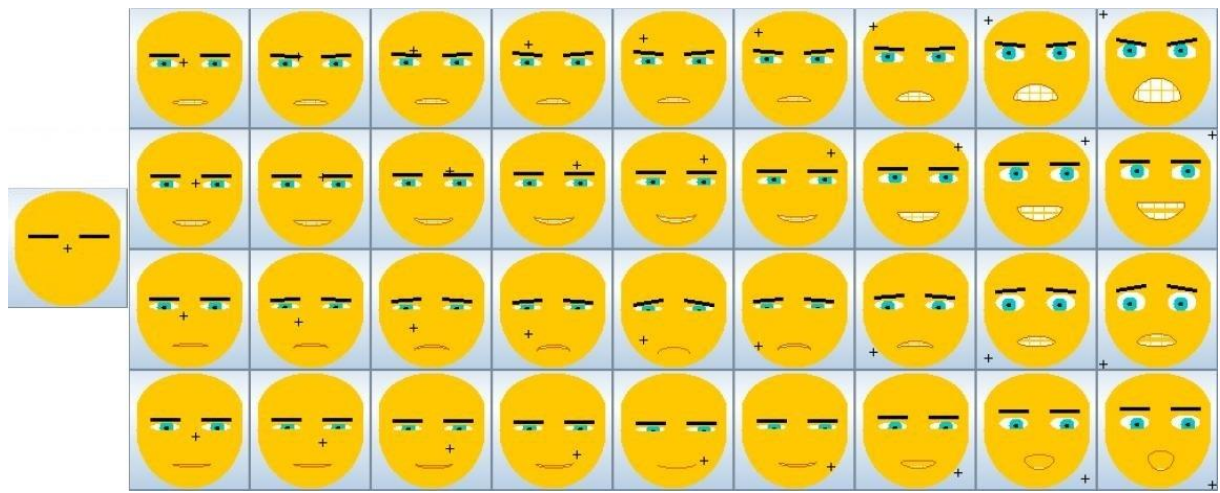

Figure 9. The AffectButton facial expressions (the black cross represents the mouse click position that selects each of the images) (Broekens and Brinkman, 2013).

The images are designed to reflect emotions on a three dimensional scale: *Pleasure* (or valence), *Arousal*, and *Dominance* (PAD) according to the emotion theory proposed in (Bradley and Lang, 1994). In this theory *P* indicates how positive/negative an emotion is; *A* indicates the level of activation of the emotion, and *D* indicates whether the environment is imposing influence over the respondent or the inverse. For instance, the difference between anger and fear can now be accounted for as: anger is a "negatively arousing dominant emotion", whereas fear is a "negatively arousing submissive emotion". The facial expressions in the AffectButton map to 36 major points in the PAD affective space. After each game session, the index of the selected AffectButton image (ranging from 0 to 36) and the corresponding values of each of the six basic emotions as reported by the player is saved into a central database server.

## 4.3.2. Study Participants

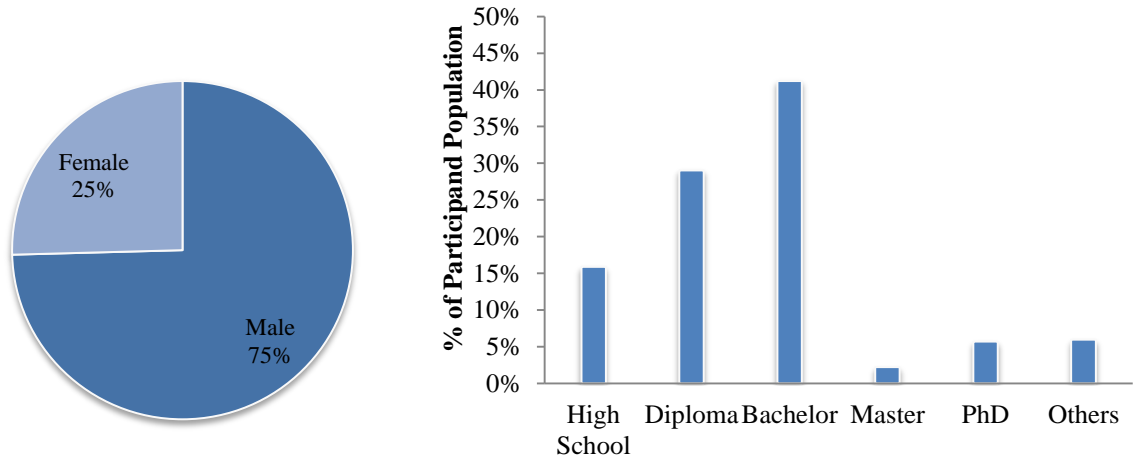

(a) Gender  (b) Highest Level of Education

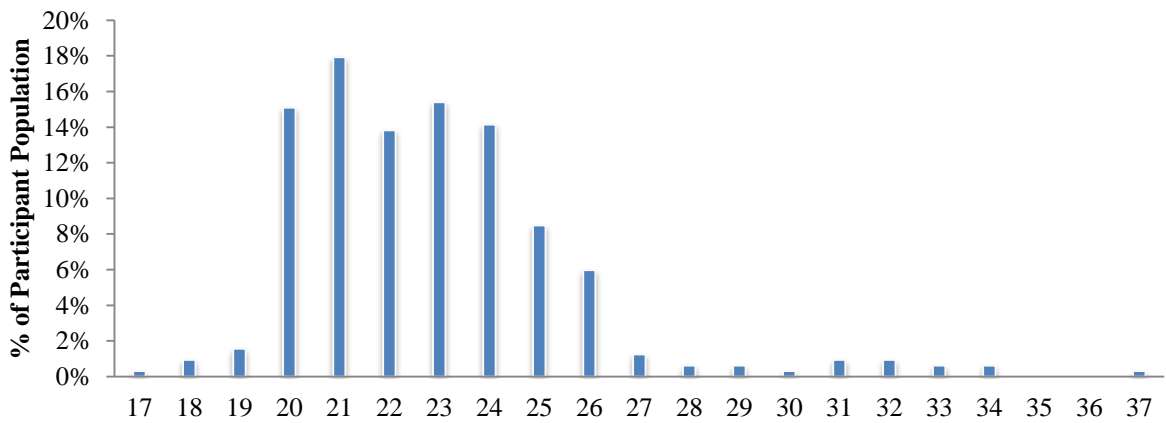

(c) Age

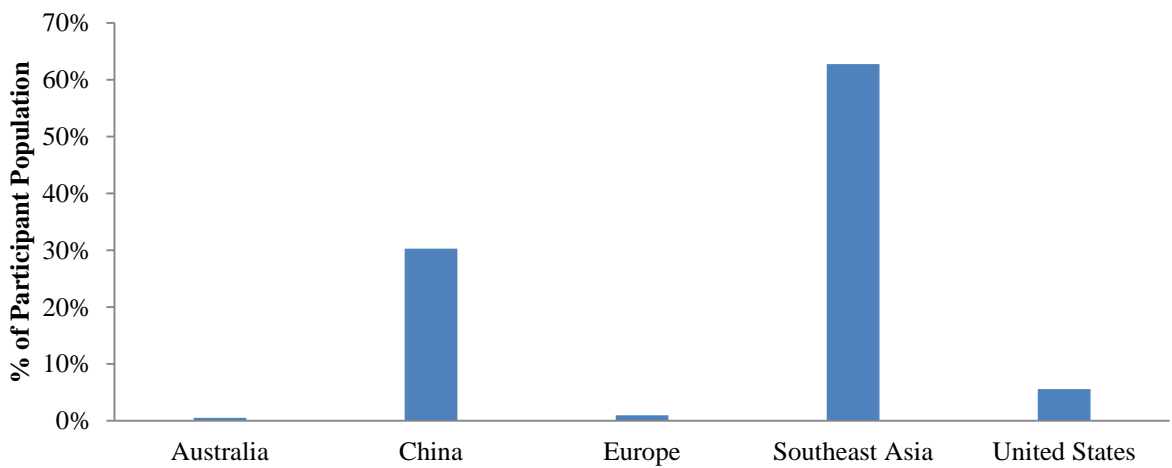

(d) Geographical distribution

Figure 10. Characteristics of the study participants.

The AM game platform has been used by Nanyang Technological University (NTU), Singapore and the Beihang University, Beijing, China as a coursework tool for undergraduate students learning software engineering. In addition, the platform has been demonstrated to multi-agent researchers during the 13th International Conference on Autonomous Agents and Multiagent Systems (AAMAS) in 2014 (Yu et al., 2014). The AM game is made available for anyone to download on a publicly accessible website from February 2014.

A total of 413 people participated in our study by downloading and playing the AM game. One in four of them are female and the rest are male (Figure 5(a)). The highest levels of education of the participants range from high school to PhD with the majority of them with tertiary level diplomas or degrees (over 70%) as shown in Figure 5(b). The participants' ages range from 17 to 37. Most of them are between the age of 20 to 25 (Figure 5(c)). In terms of geographical distribution, as shown in Figure 5(d), most participants are located in China and Southeast Asian countries (including Malaysia, Singapore, and Vietnam) with a small proportion of them located in Australia, Europe and the United States of America.

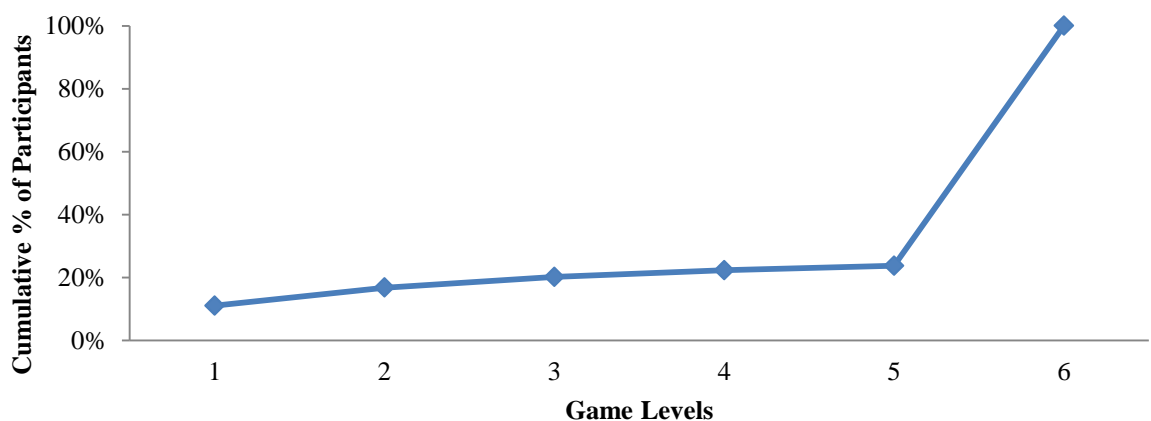

Figure 11 Statistics on game levels completed by study participants.

The cumulative percentage of participants who have completed up to Level 1 to 6 in the AM game is shown in Figure 6. It can be observed that about 80% of them finished all 6 levels of the game. The participants spent between 1 to 15 minutes per completed game session (one completed session contains all activities performed by the player for one completed game level). As illustrated in Figure 7, most participants complete a session within

Chapter 4.Empirical Studies for Building Emotional Interaction between ACA and the Users

7 minutes. We received a total of 2,267 valid emotion reports from the participants over all game sessions. In about 30% of them, the participants reported discrete emotions (i.e., only one of the six basic emotions). The remaining 70% of them denote mixed emotions. As shown in Figure 8, over half of all the reports contain mixed emotions involving all 6 basic emotions.

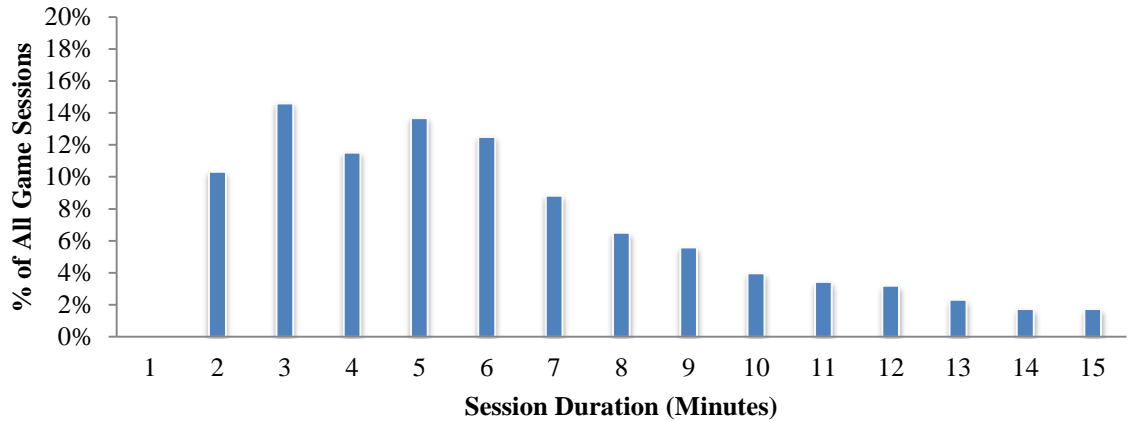

Figure 12 Distribution of game session durations.

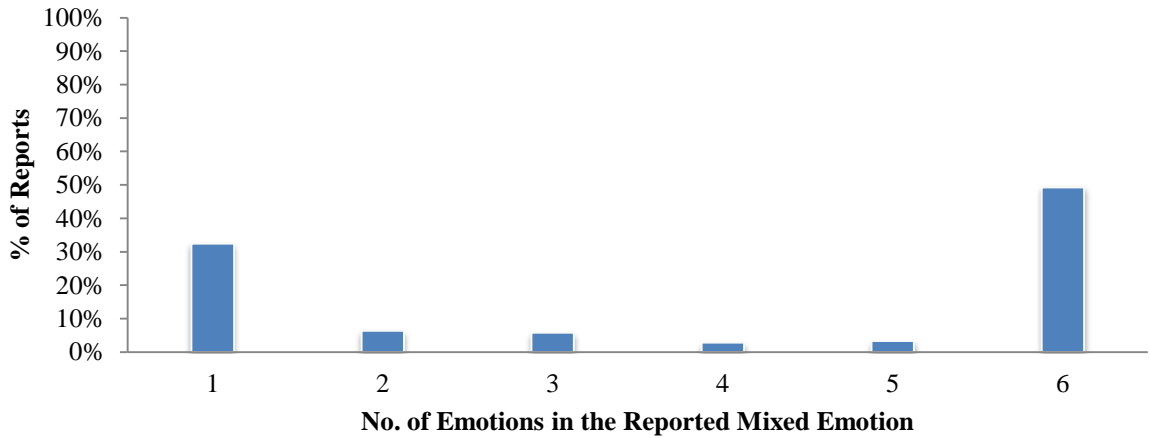

Figure 13 Discrete and Mixed Emotions reported by the study participants.

## 4.4. Results Analysis

Several emoticons were selected less than 10 times by one gender. Due to their lack of representativeness, we exclude them from the analysis. After this, the top two high scores for the emoticons selected by each gender for each emotion are shown in the following figure:

| **Emotion** | **Female** | | | **Male** | | |
|---|---|---|---|---|---|---|
| | No. | Icon | Score | No. | Icon | Score |
| **Happiness** | 15 | 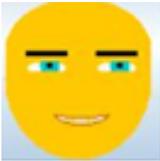 | 7.26 | 18 | 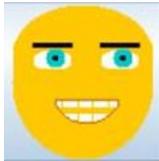 | 8.05 |
| | 18 | 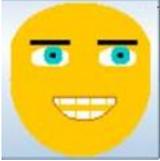 | 6.38 | 17 | 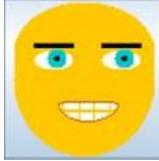 | 7.88 |
| **Excitement** | 18 | 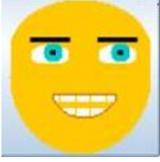 | 4.88 | 17 | 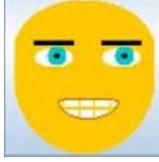 | 5.33 |
| | 12 | 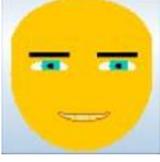 | 4.72 | 18 | 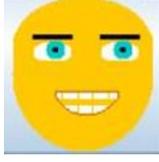 | 5.22 |

| Surprise | 18 | 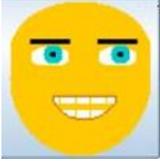 | 4.70 | 17 | 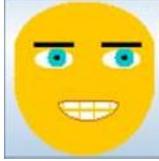 | 4.71 |
|---|---|---|---|---|---|---|
| | 12 | 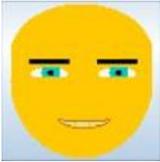 | 3.48 | 27 | 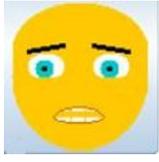 | 4.31 |
| Boredom | 12 | 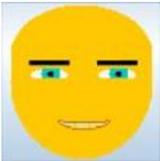 | 3.86 | 9 | 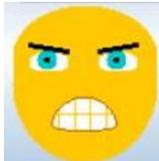 | 5.03 |
| | 3 | 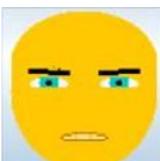 | 3.80 | 7 | 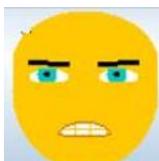 | 4.90 |
| Anger | 9 | 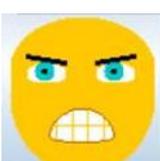 | 6.48 | 9 | 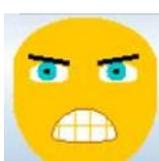 | 6.26 |
| | 18 | 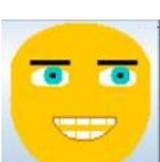 | 3.25 | 2 | 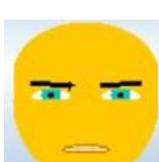 | 4.19 |

| **Sadness** | 23 | 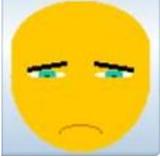 | 4.17 | 22 | 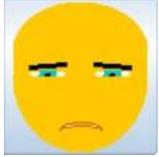 | 5.00 |
|---|---|---|---|---|---|---|
| | 18 | 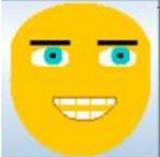 | 3.61 | 9 | 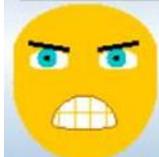 | 4.45 |

Table 2. The results of facial expression study

1.  Women are more subtle and sensitive in choosing facial expressions, especially for positive emotions. For happiness, though women agreed with men that No. 17 and 18 represent happiness. They believed No. 15 to be a happier face. The smaller eyes and mouth of this emoticon make it looks more peaceful compared to No. 17 and No. 18. The same applies in the case of "excitement". The emoticon No.12 is believed to be excitement by female participants. For male participants, it was not so representative.

2.  Emoticons No. 17 and 18 are popular among both genders. Both men and women tend to select them to represent several emotions including happiness, excitement and surprise. We can conclude that these three emotions are related to each other. However, there are also some women believing these two emoticons represent sadness and anger. This unexpected result showed significantly in both No. 17 and 18. The tendency is obvious in the following figure:

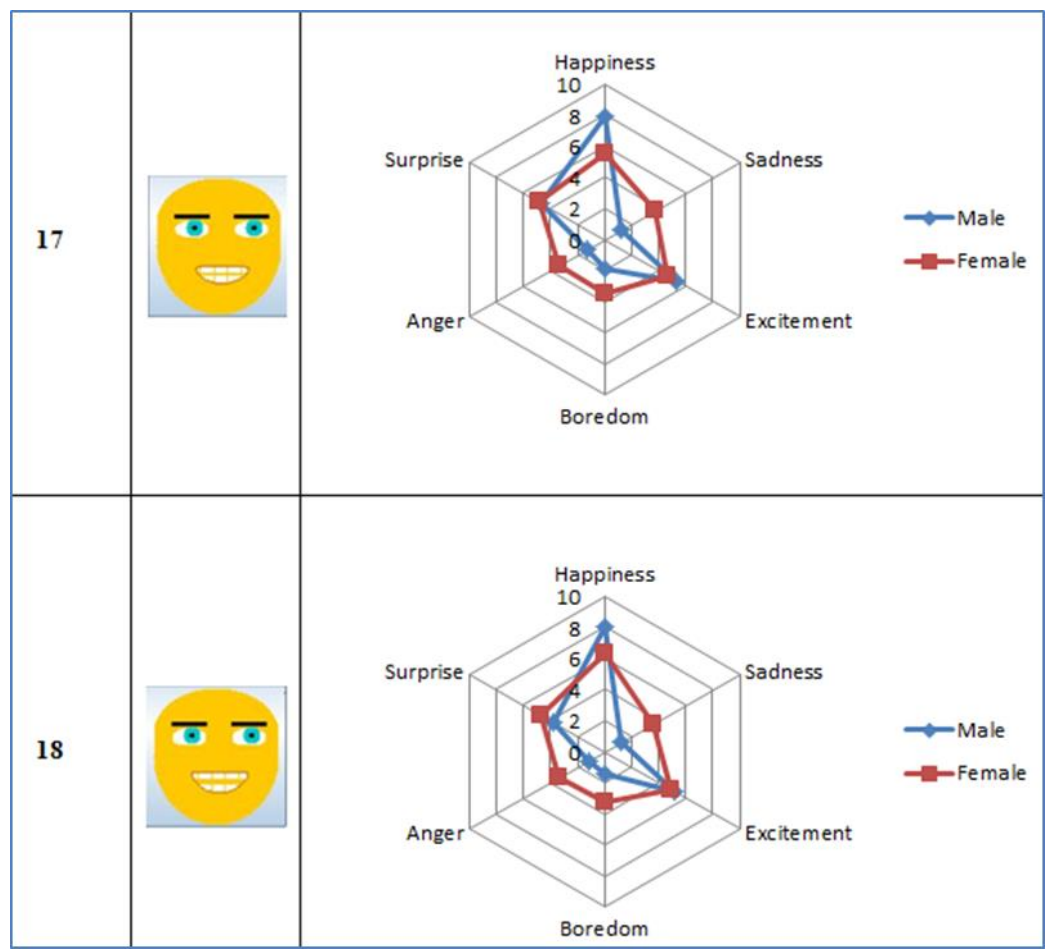

Figure 14 Different ideas about expression NO. 17 and NO. 18 between men and women

3.  For the emotion of "boredom", men give the highest score to emoticon No. 9 and 7. Both these two emoticons are related to anger. However, highest scores from female participants are given to No. 12 and 3. These two emoticons are blander than the male choices. No. 12 is also selected by many women as "happiness" and "excitement". For men, "boredom" always makes them angry. For women, "boredom" is not related much to other emotions. Any weak emotion stimulation can be boredom, even happiness or excitement.

4.  Emoticon No. 9 has shown interesting outcomes. Both male and female participants agreed that this emoticon best represents the emotion of "anger". However, men also give it a high score as 5.03 for "boredom" and 4.45 for "sadness". At the same time, women think this emoticon has little to do with "boredom" or "sadness" (with the scores of 2.00 and 2.04, respectively). We can conclude that emoticon No.9 is a complex mixed emotion expression from a male perspective.

Chapter 4.Empirical Studies for Building Emotional Interaction between ACA and the Users

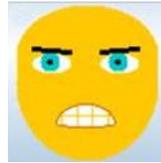

Figure 15 Facial expression NO. 9

5. The emotion of "surprise" has also shown gender differences. The first choices to represent "surprise" are similar for both genders. Men believed that emotion No. 27 also contains surprise; while for women, this emoticon represents more sadness which is also agreed by men.

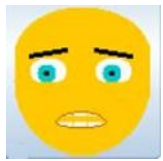

Figure 16 Facial expression NO. 27

## 4.5. Study Limitations

In this section, we discuss the study limitations based on threats to the internal and external validities as described in (Wohlin et al., 2000). We list all possible threats, measures taken to reduce the risks, and suggestions for improvements in future studies where applicable.

### 4.5.1. Internal Validity

An internal validity threat which may have affected our study is the lack of control of the following variable: the participants' skills in solving the challenges posed by the AM game. As a result, we cannot control the types of emotional responses that may be triggered as a result of playing the AM game. Nevertheless, as each participant is required to play through 6 game levels with different difficulty, the large number game sessions (over 2,200 of them) provide participants with many opportunities to experience different emotions, and decrease the probability of this threat affecting our outcomes to some extent.

### 4.5.2. External Validity

A factor that might reduce the external validity of our study is the use of students as subjects. Nevertheless, students have demonstrated that they can play an important role in experimentation in fields such as software engineering (Kitchenham et al., 2002; Lin et al., 2014). However, to be conservative, we refrain from generalizing our results to other professions or age groups.

Another threat to the external validity of our study is that the participants are predominantly of Asian origin. Previous research works have pointed to significant differences in the body gestures of people from different cultures (Endrass, 2013). As we are studying a personal and subjective topic, we suspect that the association between mixed emotions and facial expressions may be different for people from different cultures, too. Thus, we refrain from generalizing our results to people from other cultural backgrounds. In future studies, we plan to replicate the experiment with participants from more diverse backgrounds, possibly through online crowdsourcing in platforms such as the Amazon's Mechanical Turk.[1]

## 4.6. Implications

Externalizing emotions as "anger" can be recognized from 2D faces for both genders with relative ease. One good example is emoticon No. 9. Both male and female participants believe it is showing the emotion of "anger" much more strongly than others.

Women are more sensitive in recognizing facial expressions than men. They can decode subtle expressions of complex emotions. This capability has two potential applications in virtual embodied agent design. Firstly, intelligent agents can transfer emotional information to women through tiny change on their facial expressions. For men, the changes in facial expression need to be more pronounced. Secondly, women tend to explain facial expressions in a complex way. For them, different emotions, even totally opposite ones, can exist in one single facial expression. This raises a challenge to the intelligent agent design in emotional communication through facial expressions with female users. For example, we still have no idea about how women interpret sadness and anger from face No. 18. Problems as this need further studies to solve.

---

[1] https://www.mturk.com/mturk/welcome

# Chapter 5 Future Work

In Chapter 3, we proposed the role fulfilling model as a design philosophy to guide the choices of appropriate artificial companion agents for the elderly. In this chapter, we will design a serious game platform to study people's self-reported and actual behaviours under the framework of the role fulfilling model. The effectiveness of this model will also be evaluated based on this game platform. For these purposes, we need a large scale of senior users. This is the reason that we choose the serious game and crowdsourcing as experimental methods. Beside this quantitative study, a Phenomenology based qualitative study will also be conducted for understanding the reasons behind people's choices.

## 5.1. Design of the Quantitative Study

### 5.1.1. Introduction of the game

The proposed game is called Happily Aging in Place. This game simulates the daily life of a senior lady named Aunt Lily. The interface is shown in Figure 17:

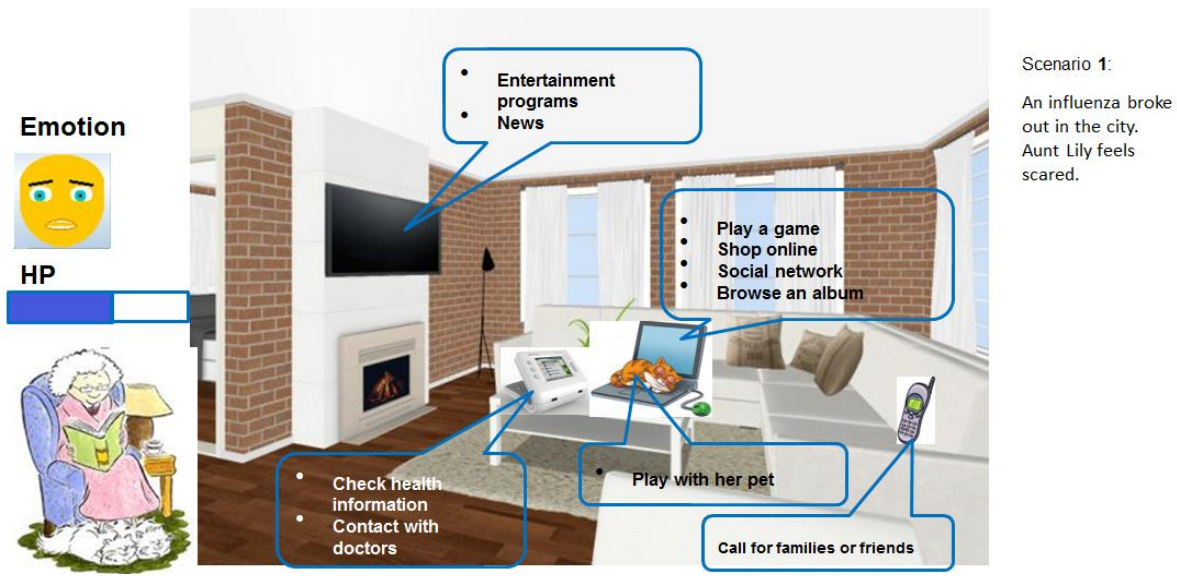

Figure 177 Interface of the game Happily Aging in Place

# Chapter 5 Future Work

This lady lives alone in her own house. In each level, the game presets a scenario and the emotion of Aunt Lily. The emotion status, which is shown as the smiley, is related to her health which is represented by the health point bar. The player's task is to moderate Aunt Lily's emotion by triggering the agents in her living room. By triggering some agents which can positively influence Aunt Lily's emotion, the health point will be enhanced; otherwise, Aunt Lily's health point will not improve. When the health point crosses a threshold of danger (which means Aunt Lily is sick), the player will fail to pass this level. When the player decides his/her companion agent selection, he/she can submit the selection. The health point of Aunt Lily at that point in time will be the final score of the player for this level. A rank of scores of all the players will be shown later. This rank aims to encourage the player to keep playing this game for higher ranks.

After the first scenario, the player will be asked two questions. The first one is "how do you determine your strategy in choosing agents". The players choose one answer from "I believe they can help", "I am interested in these agents" and "I just try randomly". Besides these, if the play has other answers, he/she can also express it by typing it out. The answer of this question will help us to understand the player's strategy and preference for artificial companion agents. For the young players, the results can help us understand what kind of devices they may buy for their parents. And for the senior players, the results can directly shows what kind of companion agents they would like to use.

The second question aims to understand the emotion of the player after the first round. The player needs to express their emotions after the game under the Positive and Negative Affect Schedule (PANAS). This 5-point scale schedule can quantify the players' emotion after the game. Higher score represents a better mood for the player.

After these two pages, the player will enter the next level. The task is the same as the first level, but under a different scenario. During this round, some of the players will be randomly allocated with a companion agent who will be shown on the right-hand side of the game interface as Figure 18 is showing.

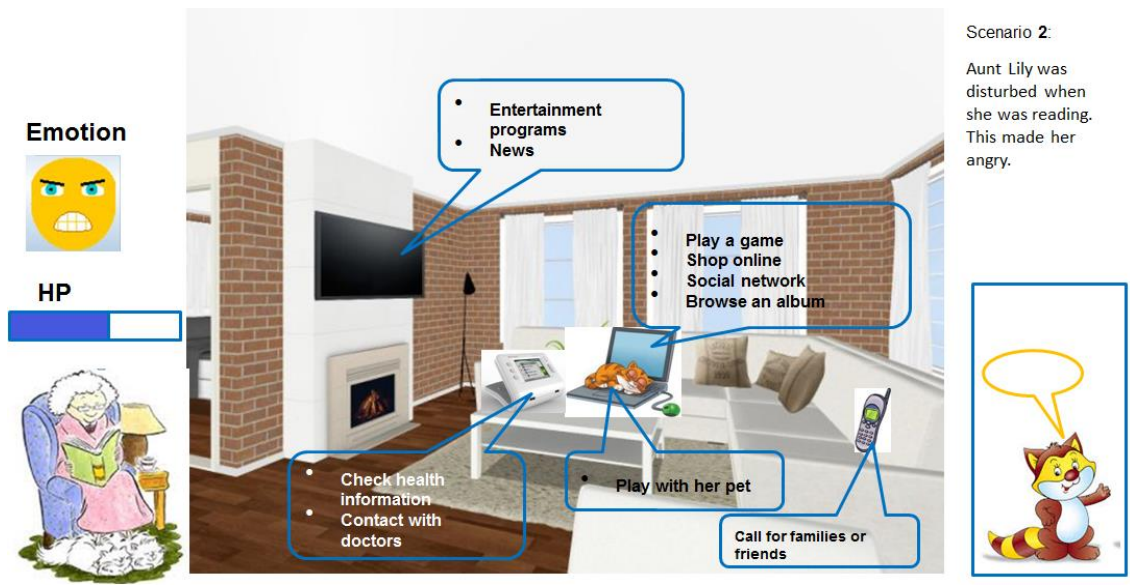

Figure 18 Interface of the game with the companion agent[2]

This agent can communicate with the player through texts, gestures, and facial expressions. Its words and movements are determined by the result of the player's emotional test after the first round. Based on this result, the system will find the appropriate functions for this agent to perform following the Role Fulfilling Model.

After scenario 2, the player's mood will be tested again with PANAS. The results from the players with the agent will be compared against the results from those without it.

### 5.1.2. Experiment design and hypothesis.

This game will be published on the Internet to leverage the power of crowdsourcing to reach a large user base. The experiment will be guided by Solomon Four Groups Test. This test is a standard pretest-post-test two-group design, and the post-test-only control group design. Group 1 and Group 2 will play the game under both scenario 1 and 2. The only difference during the processes is that Group 1 has the agent to accompany them in the second scenario. Group 3 and Group 4 play only the second scenario. Group 3 has the companion agent while Group 4 does not. The score of emotional tests from all the groups are marked as table 3:

---

[2] All the icons in this figure come from the Internet

| Randomized Groups | Pretest (After Round 1) | Agent in Round 2 | Posttest (After Round 2) |
|---|---|---|---|
| **Group 1** | $P_o1$ | With the Agent | $P_r1$ |
| **Group 2** | $P_o2$ | No Agent | $P_r2$ |
| **Group 3** | | With the Agent | Pr 3 |
| **Group 4** | | No Agent | Pr 4 |

Table 3. Marks of the emotional test results

Our hypothesis is if the result of emotional test of Group 1 improved significantly more than Group 2, the agent is proved to be effective. That means if $P_r1 - P_o1 > P_r2 - P_o2$, we can say that our agent is appropriate.

The two extra control groups 3 and 4 in this experiment can help us test whether the first round of the game itself has an effect on the subjects. The comparison between the post-test results of Groups 3 and 4, marked by Pr 3 and Pr 4, allows us to determine if the game of scenario 1 has influenced the results and to what extent.

And the comparison between the Group 2 pretest (Po2) and the Group 4 post-test (Pr 4) allows us to establish if any external factors have caused a temporal distortion. In this study, this comparison can help us avoid the effect of difference of scenarios.

## 5.2. Phenomenology Based Qualitative Study Design

Although the quantitative study may offer us a clear picture of the effectiveness of the selected agent and the Role Fulfilling Model, it cannot give us the reasons behind the observed phenomenon or information about user experience. To solve this problem, a qualitative study is necessary.

Phenomenology is the study of subjective experience (Hardy, 2001). In This book draft, we will involve at least 5 senior participants from different genders and nationalities to observe their experience in the real world. Interviews about the reason of their behaviours and strategies will be conducted during the observation. The findings from this study can mitigate the flaws of invisibility, uncertainty and unexplained observations from the human computation based user study.